\input harvmac

\def\cO{{\cal O}}
\noblackbox

\lref\MaldacenaVR{
  J.~M.~Maldacena,
  ``Non-Gaussian features of primordial fluctuations in single field inflationary models,''
JHEP {\bf 0305}, 013 (2003).
[astro-ph/0210603].
}
\lref\MaldacenaRE{
  J.~M.~Maldacena,
  ``The Large N limit of superconformal field theories and supergravity,''
Adv.\ Theor.\ Math.\ Phys.\  {\bf 2}, 231 (1998).
[hep-th/9711200].
}
\lref\LarsenPF{
  F.~Larsen and R.~McNees,
  ``Inflation and de Sitter holography,''
JHEP {\bf 0307}, 051 (2003).
[hep-th/0307026].
}

\lref\LarsenKF{
  F.~Larsen and R.~McNees,
  ``Holography, diffeomorphisms, and scaling violations in the CMB,''
JHEP {\bf 0407}, 062 (2004).
[hep-th/0402050].
}

\lref\LarsenET{
  F.~Larsen, J.~P.~van der Schaar and R.~G.~Leigh,
  ``De Sitter holography and the cosmic microwave background,''
JHEP {\bf 0204}, 047 (2002).
[hep-th/0202127].
}

\lref\LarsenPF{
  F.~Larsen and R.~McNees,
  ``Inflation and de Sitter holography,''
JHEP {\bf 0307}, 051 (2003).
[hep-th/0307026].
}

\lref\Witten{
  E.~Witten,
  ``Quantum gravity in de Sitter space,''
[hep-th/0106109].
}
\lref\HarlowKE{
  D.~Harlow and D.~Stanford,
  ``Operator Dictionaries and Wave Functions in AdS/CFT and dS/CFT,''
[arXiv:1104.2621 [hep-th]].
}
\lref\AnninosUI{
  D.~Anninos, T.~Hartman and A.~Strominger,
  ``Higher Spin Realization of the dS/CFT Correspondence,''
[arXiv:1108.5735 [hep-th]].
}
\lref\StromingerGP{
  A.~Strominger,
  ``Inflation and the dS/CFT correspondence,''
JHEP {\bf 0111}, 049 (2001).
[hep-th/0110087].
}
\lref\StromingerPN{
  A.~Strominger,
  ``The dS/CFT correspondence,''
JHEP {\bf 0110}, 034 (2001).
[hep-th/0106113].
}

\lref\OsbornCR{
  H.~Osborn and A.~C.~Petkou,
  ``Implications of conformal invariance in field theories for general dimensions,''
Annals Phys.\  {\bf 231}, 311 (1994).
[hep-th/9307010].
}
\lref\Lyth{
  D.~H.~Lyth,
  ``What would we learn by detecting a gravitational wave signal in the cosmic microwave background anisotropy?,''
Phys.\ Rev.\ Lett.\  {\bf 78}, 1861 (1997).
[hep-ph/9606387].
}

\lref\AdeXNA{
  P.~A.~R.~Ade {\it et al.}  [BICEP2 Collaboration],
  ``BICEP2 I: Detection Of B-mode Polarization at Degree Angular Scales,''
[arXiv:1403.3985 [astro-ph.CO]].
}
\lref\hif{
  J.~Garriga and Y.~Urakawa,
  ``Holographic inflation and the conservation of $\zeta$,''
[arXiv:1403.5497 [hep-th]].
}
\lref\hie{
  U.~Kol,``On the dual flow of slow-roll Inflation,''
JHEP {\bf 1401}, 017 (2014), [arXiv:1309.7344 [hep-th]].
}

\lref\hid{
  J.~Garriga and Y.~Urakawa,
  ``Inflation and deformation of conformal field theory,''
JCAP {\bf 1307}, 033 (2013).
[arXiv:1303.5997 [hep-th]].
}
\lref\hic{
  A.~Bzowski, P.~McFadden and K.~Skenderis,
  ``Holography for inflation using conformal perturbation theory,''
JHEP {\bf 1304}, 047 (2013).
[arXiv:1211.4550 [hep-th]].
}
\lref\hib{
  K.~Schalm, G.~Shiu and T.~van der Aalst,
  ``Consistency condition for inflation from (broken) conformal symmetry,''
JCAP {\bf 1303}, 005 (2013).
[arXiv:1211.2157 [hep-th]].
}
\lref\hia{
  E.~Halyo,
  ``Holographic inflation,''
JHEP {\bf 0402}, 062 (2004).
[hep-th/0203235].
}
\lref\hiz{
  A.~Bzowski, P.~McFadden and K.~Skenderis,
  ``Holographic predictions for cosmological 3-point functions,''
JHEP {\bf 1203}, 091 (2012).
[arXiv:1112.1967 [hep-th]].
}
\lref\hiy{
  P.~McFadden and K.~Skenderis,
  ``Cosmological 3-point correlators from holography,''
JCAP {\bf 1106}, 030 (2011).
[arXiv:1104.3894 [hep-th]].
}
\lref\hix{
  P.~McFadden and K.~Skenderis,
  ``Holography for Cosmology,''
Phys.\ Rev.\ D {\bf 81}, 021301 (2010).
[arXiv:0907.5542 [hep-th]].
}
\lref\Allen{
  B.~Allen,
  ``The Graviton Propagator in De Sitter Space,''
Phys.\ Rev.\ D {\bf 34}, 3670 (1986).
}

\lref\hig{
  S.~Kawai and Y.~Nakayama,
  ``Improvement of energy-momentum tensor and non-Gaussianities in holographic cosmology,''
[arXiv:1403.6220 [hep-th]].
}

\Title{\vbox{\baselineskip12pt 
}}
{\vbox{\centerline {BICEP2 and the Central Charge} \medskip \centerline{ of Holographic Inflation}} } 

\centerline{Finn
Larsen\foot{larsenf@umich.edu}
and 
Andrew Strominger\foot{andy@physics.harvard.edu}
}

\bigskip
\vskip.2cm 
\centerline{${}^1$\it{Michigan Center for
Theoretical Physics,}} \centerline{\it{University of Michigan, Ann
Arbor, MI 48109-1120, USA}}
\medskip

\centerline{${}^2$\it{Center for the Fundamental Laws of Nature, Harvard University}}\centerline{\it{ Cambridge MA,
USA}}

\baselineskip15pt

\vskip .3in

\centerline{\bf Abstract}
Holographic inflation posits that the inflationary deSitter era of our universe is approximately described by a dual  three-dimensional Euclidean CFT living on the spatial slice at the end of inflation. We point out that the BICEP2 
results determine the central charge of this putative CFT to be given by $C_T=1.2\times 10^9$. 

\Date{}

\newsec{Introduction}

String theory on anti-deSitter space (AdS) and near-AdS is known to have 
a dual description in terms of a conformal field theory (CFT) or near-CFT living on a holographic plate at the boundary of AdS \MaldacenaRE. One hopes this compelling holographic relation could apply to the real world. Unfortunately, our universe does not resemble AdS. It may  however resemble its close cousin deSitter space (dS) at both late and early times. It is  therefore of special interest  to understand holographic duality for dS. 

These considerations motivated the proposal \refs{\StromingerGP\LarsenET- \MaldacenaVR} (recent discussions include \refs{ \hix \hiy \hiy \hib \hic \hid \hie \hif -\hig}) that the inflationary era of our universe has a dual description in terms of a 3D near-CFT$_3$ living on a Euclidean holographic plate stationed at the end of inflation.  dS-invariance of the inflationary geometry is equivalent to conformal invariance of the CFT$_3$. Every 4D bulk field in the inflationary era is dual to an operator in this CFT$_3$, and correlation functions of bulk fields are given by CFT$_3$ correlation function of the dual operators. These correlation functions are in fact what is measured in the sky, so the astronomical data is close to that which naturally characterizes the CFT$_3$.  

Every CFT$_3$ has a traceless conserved three-dimensional stress tensor $T_{ij}$. The two point function of this stress tensor is fixed by conformal symmetry to be \OsbornCR 
\eqn\da{
\langle T_{ij} (\vec x) T_{kl}(0)\rangle = {C_T\over |\vec x|^6} I_{ij,kl}(\vec x)~,
}
with
\eqn\db{
I_{ij,kl} = {1\over 2} \left( I_{ik} I_{jl}+ I_{il} I_{jk}\right)-
{1\over 3} \delta_{ij} \delta_{kl}~.
~~~~~~
I_{ij}(\vec x) = \delta_{ij} - 2 {x_i x_j \over x^2}~.
}
The number $C_T$ here is referred to as the central charge and characterizes the number of degrees of freedom of the CFT$_3$.\foot{For a theory with $N$ free scalar fields $C_T = 3N/32\pi^2$.}

The main purpose of this paper is simply to point out that BICEP2 has made a direct measurement of the central charge 
$C_T$  for the CFT$_3$ dual of inflation. This comes about because the holographic dictionary relates correlators of the 4D metric to correlators of the CFT$_3$ stress tensor. We will find that 
\eqn\mv{C_T=1.2\times 10^9~.} 
In previous work,  data on the scalar mode fluctuations and their tilt was translated into  various CFT$_3$ operator dimensions and beta functions  \refs{\LarsenET,\hia, \LarsenPF} and absence of non-gaussianities used to bound three-point functions \MaldacenaVR.  BICEP2  now provides us with important information about a central number characterizing the CFT. 

While there has been good recent progress on 4D toy models \AnninosUI,  there is no complete detailed proposal on the table for what this CFT$_3$ might be in the real universe: rather its properties are deduced from measurement.  Knowing the complete CFT$_3$ would be tantamount to knowing every detail of inflation. This is not, in our opinion,  something we can realistically  strive for. However in the absence of such a complete description, the existence of a holographic description of inflation is practically a tautology: nearly any collection of equal-time dS-invariant  correlation functions can be defined to be those of some CFT$_3$.\foot{This was spelled out in a Hamilton-Jacobi formalism in \LarsenET. Note that one does not demand the usual 'good' properties of the CFT$_3$ such as unitarity. Indeed it is known that the CFT$_3$ cannot be unitary \StromingerPN. Demanding  special properties of the CFT$_3$ would certainly constrain the structure of inflation.  }
This raises the question: What is holographic inflation good for on the observational front? Here are several possible answers:

\itemitem{$\bullet$}
Nothing.

\itemitem{$\bullet$}
The data characterizing a particular CFT$_3$ - central charge, operator dimensions, etc. - is very different than that characterizing a particular model of inflation - the number of inflatons,  their potentials, etc. This reorganization of the data could lead to conceptual insights. In particular, what is natural in one formulation is not necessarily natural in another and vice versa \StromingerGP. The reorganization may also be useful simply as a computational strategy,  for example the analyses of  non-gaussianities in \MaldacenaVR\  or relations among inflationary parameters \LarsenET. 

\itemitem{$\bullet$}
All ordinary models of inflation assume the existence of a local  quantum field theory weakly coupled to gravity and some kind of inflaton. It is not obvious that this need to be the case. Indeed the proximity of the inflation scale to the Planck or string scales raises the possibility that  the scales may not really be separated.  Moreover the  necessity of an inflaton whose shift exceeds the Planck  scale \Lyth\ seems problematic. Holographic inflation assumes only the symmetries, and not the existence of a weakly-coupled quantum field theory local on scales much shorter than the dS radius.  In this sense it is more general than the usual inflationary models, while still fully capturing the symmetry (breaking) structure largely responsible for the phenomenological success of those models. 

In section 2 we recap the basic result of BICEP2. In section 3 we describe holographic inflation in a little more detail, in particular the precise relation between 4D metric correlators 
and 3D stress tensor correlators,  and derive the result for the central charge. 

\newsec{Measuring quantum correlations of the metric}

During the inflationary era the spacetime geometry is well approximated by the dS metric
\eqn\dsl{ds^2=-dt^2+e^{2Ht}\delta_{ij}dx^i dx^j.} For $H\ll { 1 \over t_p}$, where $t_p =\sqrt{8\pi G\hbar}$ is the Planck time (in units where $c=1$), quantum fluctuations of the metric are small and the free field approximation can be applied.  We take the end of inflation to be at $t=0$. We are interested in the transverse  traceless modes 
\eqn\fg{ \delta^{T}g_{ij}\equiv\psi_{ij}~, ~~~~ \nabla^i \psi_{ij}=\delta^{ij}\psi_{ij}=0~,}
which correspond to physical gravity waves. It is conventional and convenient to transform to spatial momenta $\vec k$,
\eqn\ft{\psi_{ij}(\vec x)=\int {d^3 k \over (2\pi)^3}e^{i \vec k \cdot \vec x}\psi_{ij}(\vec k)~,}
where the time argument is fixed to $t=0$. 
 We introduce a complex polarization vector $e_i$ 
such that 
\eqn\pv{e_ik^i=0~,~~~\delta^{ij}e_ie_j=0~, ~~~\delta^{ij}e_i\bar e_j=1~,}
where bar here and herafter denotes complex conjugation. 
We may then expand
\eqn\rxa{\psi_{ij}(\vec k)=\psi(\vec k)e_ie_j+\bar \psi (\vec k)\bar e_i \bar e_j~.}
The single complex scalar $\psi$ represents the two polarization modes of the graviton. The
 two point function of $\psi_{ij}$ is determined by the $SO(4,1)$ dS invariance together with the demand that it have the standard short-distance singularity \Allen
\eqn\ac{
 \langle \psi (\vec{k})\bar  \psi (\vec{k'}) \rangle = (2\pi)^3 \delta(\vec{k}-\vec{k'}) {1\over 2} P_t(k) ~,
}
with the power spectrum
\eqn\acf{
P_t (k) = {4 (Ht_p)^2\over |k|^3}\equiv{2\pi^2 \over |k|^3}A_t ~,
}
where the dimensionless amplitude is
\eqn\rto{A_t=2\left( {Ht_p\over  \pi}\right)^2~.}

These correlated, polarized metric fluctuations are present at the end of inflation and lead to correlated polarizations in the CMB, with amplitudes proportional to $P_t$. Recently, the latter correlations were directly measured \AdeXNA\ and found to imply via \acf\ 
\eqn\yol{A_t=5\times 10^{-10}.}
Equivalently, the Hubble constant of inflation is 
\eqn\jsr{Ht_p=5\times 10^{-5}.}

\newsec{Holographic Cosmology}
A variety of theoretical investigations have suggested that a consistent quantum theory of gravity in a spacetime $M$ should have a fully equivalent ``holographically dual'' description in terms of a quantum field theory which lives on a ``holographic plate'' at the boundary of $M$. This is known to be true in some special string theoretic examples, but the more general situations, including the cosmological context, are very far from understood. 
In this section we will give a nontechnical description of the holographic approach to cosmology, and present a few essential formulae. We refer the reader to \HarlowKE\ for a cogent recent discussion of the holographic dictionary for dS, and to \LarsenKF \foot{In addition to the formula for the central charge given herein, these authors relate the spectral tilt and slow-roll expansion parameters to operator dimensions and beta-functions of the dual field theory.} where this dictionary was fully developed in the context of inflation.

We begin with the simplest case of pure dS. 
The holographic description of positively curved dS \refs{\StromingerPN,\Witten} was inspired by that of its negatively curved cousin, AdS. In certain string 
theoretic examples, quantum gravity in AdS is known to have a holographically dual description in terms of a CFT living on 
a ``holographic  plate'' at its  boundary at spatial infinity. The conformal symmetries of the CFT are dual to the isometries of AdS. 
The radial direction in AdS is holographically emergent and corresponds to RG flow in the boundary CFT. 

 How would this work in dS?  This simplest guess is that the holographic plate lives on the spatial $R^3$ slice at future infinity, and time is the holographically emergent direction. This makes sense because the isometries of dS act as the conformal group on its future $R^3$ boundary, and because the holographically emergent direction should be the expanding one. 
Initial discussions of dS holography were plagued by the absence of any concrete toy examples to flesh it out.  However recently \AnninosUI\ a solvable higher-spin example of 
4D dS holography  has been constructed, so we at least know that dS holography is mathematically self-consistent.   

Holographic dualities involving near-AdS spacetimes have also been understood in string theory. 
The structure is similar to exact AdS, the main difference being that the quantum field theory that lives on the boundary is nearly but not exactly conformally invariant. Since the inflationary era is nearly but not exactly dS (with the deviation measured by the spectral tilt), the dual theory must be a near-CFT$_3$. We can think of it as a CFT$_3$ perturbed by nearly marginal operators. 
Since the inflationary era is nearly $SO(4,1)$ invariant, the quantum field theory dual to the inflationary era should be nearly conformally invariant. The holographic plate is  taken to be the end of inflation.

The holographic dictionary relating 4D bulk dS quantities to 3D boundary CFT$_3$ quantities can be stated in a variety of formalisms. A particularly elegant one \refs{\MaldacenaVR,\HarlowKE} begins with the  Wheeler-deWitt wave function of the universe 
$\Psi[\Phi_\alpha(\vec x))],$
where $\Phi_\alpha$ denotes  all the fields such as the metric, inflaton and matter fields on a spatial slice which will be chosen to be holographic plate $t=0$ at the end of inflation.
A particular ``Hartle-Hawking" wave function $\Psi_{HH}$ is schematically defined by a path integral 
\eqn\ba{
\Psi_{HH}[\Phi_\alpha(\vec x)]  = \int {\cal D}[\Phi_\alpha(\vec x, t)] e^{iS[\Phi_\alpha]}~.  
}
On the right hand side, ingoing Hartle-Hawking boundary conditions are taken in the far past, while at $t=0$, $\Phi_\alpha(\vec x,0)$ is constrained to take the value $\Phi_\alpha(\vec x)$. Of course expressions like this are notoriously difficult to define for many reasons. However here we are interested only in quadratic metric fluctuations about dS so these problems will not bother us. 
For every field $\Phi_\alpha$, there is a dual operator $\cO^\alpha$ in the CFT$_3$. 
The dictionary is then simply that the Hartle-Hawking  wave function is the generating function of the CFT$_3$ correlators, i.e. 
\eqn\wd{\Psi_{HH}[\Phi_\alpha(\vec x)]=\langle e^{ \sum_\alpha \int d^3x \Phi_\alpha\cO^\alpha}\rangle~.} 

Let us now evaluate the explicit expression for stress tensor correlator. This CFT$_3$ operator is dual to the metric. With the standard 3D normalization
\eqn\be{
T_{ij} =  - {2\over\sqrt{-g}} {\delta S_3\over \delta g_{ij}}~,
}
 the right hand side of \wd\ becomes 
\eqn\gt{\langle e^{-{1\over 2}\int d^3x \psi_{ij}T^{ij}}\rangle = \langle e^{-{1\over 2}\int d^3x (\psi\bar T+\bar \psi T)} \rangle = \langle e^{-{1\over 2(2\pi)^3}\int d^3k (\psi\bar T+\bar \psi T)} \rangle~. }
The arguments of the last integrand are in momentum space where
\eqn\xhj{T(\vec k)=T_{ij}(\vec k)e^ie^j~,} 
and we have used the metric components defined in \rxa.

To evaluate the left hand side of \wd\ we use the formula \ac\ for the equal-time  graviton propagator in dS
\eqn\bc{
\langle\psi(\vec{k}) \bar\psi(\vec{k'})\rangle =  {\int D[\psi] \psi(\vec{k}) \bar\psi(\vec{k'}) ~|\Psi[\psi]|^2  \over
\int D[\psi]  ~|\Psi[\psi]|^2} =  (2\pi)^3 \delta(\vec{k}-\vec{k'}) {1\over 2} P_t(k)~,}
using the notation introduced in \ac. It follows that 
\eqn\bb{
\Psi_{HH}[\psi]  \sim e^{ -\int {d^3k\over  (2\pi)^3 P_t(k) }  \psi (\vec{k})  \bar\psi(\vec{k}) }\sim Z[\psi]~, 
}
up to $\psi$-independent normalizations which will cancel. 
Since $Z[\psi]$ is the generating function of the $T$ correlators in the CFT$_3$ we have
\eqn\al{\langle T(\vec k)\bar T(\vec k')\rangle={4(2\pi)^6\over Z}{\delta^2 Z[\psi]\over \delta \bar \psi(\vec k)\delta \psi(\vec k')}=(2\pi)^3 \delta(\vec{k}-\vec{k'})   {4\over P_t(k) }~.}
Transforming back to position space and restoring the indices on $T$ gives 
\eqn\da{
\langle T_{ij} (\vec x) T_{kl}(0)\rangle = {C_T\over |\vec x|^6} I_{ij,kl}(\vec x)~,
}
with $I_{ij,kl}$ given in \db\ and 
\eqn\iya{C_T= {48\over \pi^4 A_t }={24\over (\pi H t_p)^2 } ~.}
The measured value of $A_t$ \yol\ then gives the value \mv\ of the central charge as claimed in the introduction.

Applying the formula \iya\ to our future dominated by dark energy we find $C_T = 10^{121}$. The enormous value of this central 
charge is the manifestation of the cosmological constant problem in the holographic description of the universe.

\medskip

\centerline{\bf{Acknowledgements}}
We are grateful to A. Maloney for useful conversations. We are also grateful to S. Kawai and Y. Nakayama for bringing to our attention inaccurate factors of $2$ in the first version of this manuscript. We thank the Solvay Institute for hospitality while this work was initiated. This work of FL was supported in part by  DoE grant DE-FG02-95ER40899 and AS by DoE grant DE-FG02-91ER40654. 

\listrefs

\end